# The Search for Extraterrestrial Civilizations:
## A Scientific, Technical, Political, Social, and Cultural Adventure

K. I. Kellermann
National Radio Astronomy Observatory
520 Edgemont Road
Charlottesville, VA, 22901
USA
Email: kkellerm@nrao.edu

> The probability of success is difficult to estimate, but if
> we never search, the chance of success is zero.
> Cocconi and Morrison (1959)

## Abstract

I review the scientific and technical history of the Search for Extraterrestrial Intelligence (SETI), discuss the impact of the political involvement, and speculate on the nature of a successful detection and its potential social and cultural impact. Emphasis is on the development of SETI in the United States and the complementary progress in the Former Soviet Union.

**Keywords**: SETI, CETI, NASA, Ozma, Water Hole

## 1. Introduction and Disclaimer

I have only made one SETI investigation in my career. That was a short observation more than half a century ago to look for the notch in the spectrum of the unusual radio source PKS 1934-63 of the type predicted by Nicolai Kardashev (1964). It was unsuccessful, and I buried the results in a few sentences in a scientific publication that primarily discussed the astrophysics of PKS 1934-63 (Kellermann 1966). This was the first mention of a modern search for extraterrestrial intellegence in a peer reviewed scientific publication.[1] Discouraged by my lack of success, I have not spent any more time or effort in searching for other intelligent civilizations, although I note that in spite of the enormous growth in the number of serious people involved in SETI, in the increase in technical capability, and the hundreds of subsequent investigations, no one has had more success than my modest 1964 search. My interest in SETI, however, has continued unabated. Over the past half century, I have participated in many SETI conferences and workshops, and was privileged to know and learn from nearly all of the early SETI pioneers,

## 2. Background

It has now been more than 60 years since Frank Drake's pioneering Project Ozma and the first modern attempt to detect radio signals from an extraterrestrial intelligent civilization. Since that time, the size of radio telescopes used for SETI has increased from 25 m to the order of 100 meters, system temperatures have decreased by more than an order of magnitude, and the number of simultaneous frequency channels searched has increased from one to about ten billion. But sadly, the number of successful detections remains unchanged – zero.

Drake's choice of 21 centimeters for Project Ozma was based only on the availability of a 21 cm receiver and feed that was being built in Green Bank for HI observations, specifically an attempt to detect Zeeman splitting of the 21 cm hyperfine structure line (Drake 1961, 1979, 1986). While the receiver was being built, Cornell physicists Giuseppi Cocconi and Philip Morrison published their famous paper in *Nature,* calling attention the possibility of communicating with an advanced extraterrestrial civilization. Cocconi and Morrison (1959) argued that the 21 cm line was the optimum place to look because hydrogen is the most common element in the Universe, and any extraterrestrial would recognize 1420 MHz as a special frequency for interstellar communications. Perhaps this would make sense if all the extraterrestrials were theoreticians, but if they were radio astronomers, they would understand that 1420 MHz is the worst frequency to use, partly because of the radiation and absorption at this frequency from galactic hydrogen and also because any intelligent civilization would protect 1420 MHz for radio astronomy. Just as here on Earth, surely the Galactic Telecommunications Union would prohibit any radio transmissions in a band around 1420 MHz.

Following the discovery of the interstellar 18 cm OH lines in 1963 by Weinreb et al. (1963), Barney Oliver and others argued that the so called "water-hole" between 18 and 21 cm ($H+OH=H_2O$) was the obvious place that an extraterrestrial water-based civilization would transmit, just as traditional water-holes were the place for neighbors to meet. H I and OH were the only two spectral lines known at the time in the radio spectrum, and the region between 1.4 and 1.7 GHz is in the quietest part of the microwave spectrum. So there were scientific, technical, and philosophical arguments that the frequency range from 1.4 to 1.7 GHz was the optimum place for SETI searches (Oliver and Billingham 1973).

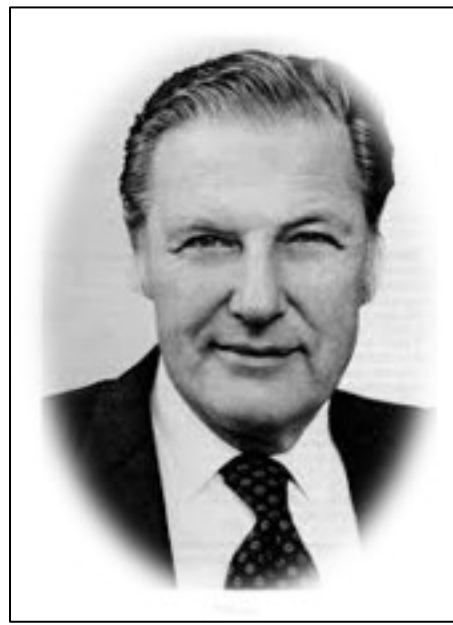

However, Charles Townes (1983, Schwartz and Townes 1961), the co-inventor of laser, suggested that instead of looking in the radio spectrum, we should look instead for infrared beacons from powerful lasers which would allow more highly focused signals and mitigate against needing to correct for Doppler shifts due to the unknown relative motion of the Earth and alien planet. Later, Paul Horwitz, at Harvard initiated an optical SETI search (Howard et al. 2004). Freeman Dyson (1967) has suggested that instead of looking for electromagnetic transmissions, we look for signs of alien engineering artifacts, or technosignatures, such as infrared radiation from their technological waste. Jay Pasachoff and Marc Kutner (1979) pointed out the advantage of using neutrinos for interstellar communications, as, unlike radio or light waves, neutrinos pass through the Galaxy without attenuation. Unfortunately, they also pass through our detectors without detection. Harris (1986) has suggested that the matter-antimatter annihilation in alien spacecraft propulsion systems could produce observable gamma rays. Ron Bracewell (1960) made the innovative suggestion that alien civilizations will send self-replicating probes to other solar systems rather than signaling from their home planet, thus facilitating practical two way communication without the long time delay that would be involved in communicating between stellar systems. Harvard astronomers Lingam and Loeb (2017) have suggested that Fast Radio Bursts may be beams used to power alien intergalactic light sails. In his highly publicized book, *Extraterrestrial: The First Sign of Intelligent Life Beyond Earth,* Avi Loeb (2021) has suggested that the interstellar asteroid Oumuamua is an alien artifact. More recently, Michael Hippke (2021) has proposed that future searches should target quantum communications due to its advantage over what he calls classical communications. I am not aware that anyone has yet suggested using gravity waves to search for extraterrestrials, but they will.

Figure 1. Barney Oliver, Hewlett Packard Vice President for Research, early SETI pioneer, originator of the "water-hole" concept, and leader of the 1971 Project Cyclops summer study.

While most searches by Western investigators have looked for narrow band signals, the early SETI studies in the USSR concentrated their efforts on looking for broad-band pulses. (Troitsky et al. 1973, 1979, Gindilis et al. 1979). To discriminate against terrestrial interference, some of these experiments used spaced antennas to search for dispersed signals resulting from propagation delays in the interstellar medium. One can only speculate on their reaction if they had had sufficient sensitivity to detect pulsars or Fast Radio Bursts. In a particularly bold experiment, Soviet radio astronomers attempted to locate a Bracewell type probe at the Earth-Moon Lagrangian points using a powerful 9.3 MHz radar facility in Gorki (Gindilis and Gurvits 2019).

The plethora of new ideas about searching for extraterrestrials has changed rapidly, with new concepts popping up on a time scale of a decade or less, yet we continue to speculate how civilizations far advanced by hundreds or thousands of years or more from our own will attempt to communicate. It is sobering to reflect on how traditional radio astronomy has changed since project Ozma 60 years ago: we haven't had a very good record in predicting new astrophysical phenomena. Quasars, pulsars, FRBs, interstellar masers, Giant Molecular Clouds, superluminal (faster-than-light) motion, the cosmic microwave background, and dark matter were all discovered during the past century ─ all unpredicted and discovered by accident as a direct result of newly developed technology. Even close at home in the solar system, radio bursts from the Sun and Jupiter, the Greenhouse effect on Venus, and the rotation of Mercury were unknown until revealed by new radio technologies. At the same time, radio astronomy has gone from arcminute resolution to an unpredicted microarcsecond resolution. Why then do we think we can understand the technology, not to mention the sociology and motivation of civilizations many hundreds and thousands of years (or more) ahead of us? Although there had been a lot of discussion about where to search for extrasolar planets, surprisingly the first extrasolar planets were found, not around a solar type star where SETI investigators were concentrating their efforts, but unexpectedly around a pulsar (Wolszcan and Frail 1992) ─ although the Nobel Prize committee apparently didn't appreciate the importance of this discovery.

One argument in favor of continuing to concentrate on radio searches is that radio transmission was the first technology to develop on Earth that was capable of interstellar communication. So even a very advanced civilization might recognize radio transmissions as the best way to communicate with our more primitive civilization, assuming that their technology developed along the same lines as ours. However, as Townes (1983) pointed out, if lasers had been invented before vacuum tubes and radio technology, our history of SETI on Earth might well have developed along different lines.

Perhaps the most ambitious attempt to understand the nature of advanced alien civilizations was the classical paper by Nickolai Kardashev (1964). Extrapolating from our past like most SETI researchers, Kardashev implicitly assumed that alien societies would continue to advance their technological development. He recognized three levels of achievement that alien societies might reach.

- Type I civilizations like our own, are able to harness energy available from their sun, as well as geothermal and tectonic energy and consume energy at a rate of about $4 \times 10^{12}$ Watts.
- Type II civilizations have harnessed the power of their sun, and consume about $4 \times 10^{26}$ Watts
- Type III civilizations have controlled the power of their galaxy at a level about $4 \times 10^{37}$ Watt.

Although Kardashev considered highly advanced civilizations with resources up to $10^{25}$ times more advanced than our own, he implicitly assumed that radio transmissions would remain the optimum method of interstellar communication. He specifically drew attention to the peaked spectrum radio sources CTA 21 and CTA 102 (Kellermann et al. 1962) which he noted had spectral energy distributions close to what he predicted was the spectral distribution that would be used by alien transmissions to optimize the information rate.

### 3. SETI Conferences, Meetings, and Workshops

Surely no other scientific topic devoid of any positive results, has been the subject of more meetings and conferences than SETI. Shortly after Project Ozma, Frank Drake and NRAO Director Otto Struve convened a small conference in Green Bank to consider the following questions:

a) What are the conditions under which intelligent radio transmissions are likely to be observable?
b) Is it worthwhile to observe with existing equipment, or are prospects for success too small to be of interest?
c) What observations are needed to make negative results interesting?

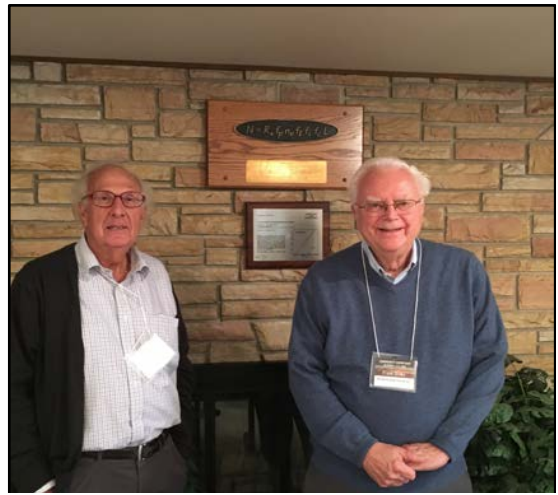

Sixty years later, these same questions still form the basis of SETI research and conferences. Only about ten people attended the 1961 by-invitation-only conference. To organize the discussion, Drake devised his now famous Drake equation. At the time, many of the parameters were unknown. Today we have a much better idea of the likelihood that stars will contain a planetary system, that some of the planets will be earthlike, and that some will lie in the Goldilocks zone ─ not too hot and not too cold but at a temperature that could support life. The big uncertainty 60 years ago was, and continues to be, $L$ in Drake's equation ─ what is the lifetime of intelligent technological civilizations? Do they go on to develop forever? Or, as debated during Cold War times, will civilizations destroy themselves by nuclear war?  Or will civilizations perish when they overuse their natural resources?  Today a pessimist might speculate that the human race will be decimated by a global pandemic scale disease.

Figure 2. Frank Drake and the author taken in July 2019 during an astrobiology conference. The setting is the lounge where Drake first presented his 1961 Drake equation, shown on the memorial plaque in the background.

Reportedly, in 1950 when discussing UFOs and considering the probability that there are many extraterrestrial intelligent civilizations in the Galaxy, Enrico Fermi famously asked, "Where is everybody?"  Generations of SETI scientists have contemplated what has become known as the "Fermi Paradox." Stephen Webb (2015) has written a provocative book with 75 possible explanations, but as our understanding about the formation of extrasolar planetary systems and the existence of earth-like planets increases, speculations have become more focused around the value of *L*.  As Carl Sagan kept reminding us, however, we must not forget Martin Rees' caution that, "The absence of evidence is not evidence of absence." (Oliver and Billingham 1973, p. 3)

Twenty-five years after the Green Bank meeting, the Ozma participants as well as other SETI researchers gathered in Green Bank to review progress and discuss strategy (Kellermann and Seielstad 1986).  The next Green Bank SETI meeting on fiftieth anniversary of Projet Ozma, paid more attention to the social, moral, legal, and religious impact of success, as well as the implications of failure.[2]

Probably the first large meeting to discuss how to communicate with extraterrestrial intelligent civilizations was organized by Yuri Parijisky in Byurakan in Soviet Armenia in 1964 (Tovmasian 1964). Like SETI meetings held over the next half century, the participants debated the multiplicity of inhabited worlds, the possible levels of development obtained by alien civilizations, the optimum techniques for establishing communication, how to distinguish RFI from legitimate signals, as well as the logistics of establishing a common language for communication. In 1971 the US and Soviet Academies of Science jointly organized the first ever international conference devoted to the search for extraterrestrial intelligent life.  Also held in Byurakan at the invitation of Viktor Ambartsumian, the conference was attended by a broad spectrum of scientists, historians, and engineers who debated related issues of anthropology, language, world history, the formation of planetary systems, the origins of life, as well as the technical issues that constrain interstellar communications (Ambartsumian 1972, Sagan 1973).  Unsurprisingly, the conference participants endorsed the importance of SETI research and the need to initiate searches, which could be modest at the start but, they suggested, might ultimately become comparable to the level of funding devoted to space and nuclear research.  At this meeting, a small group including Drake, Kardashev, Iosef Shklovsky, and Carl Sagan, inspired by the plentiful supply of Armenian cognac, agreed that before we can communicate with extraterrestrials we need to find them, so they changed the name of the field from CETI (Communication with Extraterrestrial Intelligence) to SETI (Search for Extraterrestrial Intelligence), the name which rapidly became part of the professional, political, and popular lexicon.

US and Soviet SETI researchers continued to meet on decade time scales with a 1981 conference in Tallin, Estonia (Sullivan 1982) and a 1991 conference in Santa Cruz, California (Shostack 1993) just before the fall of the Soviet Union.  SETI achieved a new level of recognition by the international community in 1979 when the International Astronomical Union convened a Joint session on SETI at its 1979 General Assembly held in Montreal, Canada.  Unlike the previous (and mostly subsequent) SETI conferences, the IAU Joint Discussion had, for the first time, a broad participation of scientists with no previous involvement in SETI.

**4.  False Alarms**

During the first 10 or 15 years after Project Ozma, there were a number of investigations to search for radio signals from an extraterrestrial civilization, almost exclusively in the US and USSR. Several investigators were allowed to use the Green Bank 140 and 300 foot telescopes for low key SETI programs, with the proviso that any results be published in the normal scientific literature and not given undue publicity (e.g., Verschuur 1973, Zuckerman and Tarter 1980, Tarter 1980). Although there were no confirmed detections of any extraterrestrial civilization, that hasn't constrained the community from writing hundreds of thoughtful original papers, reviews, and popular articles.[1]  A few investigations stand out for their apparent, albeit temporary and false, indications of a real SETI signal.

### 4.1 Project Ozma

When Frank Drake first turned the 85 foot telescope toward Epsilon Eridani, there was a strong pulsating signal, every 8 seconds, so loud it drove the chart recorder needle off scale and Drake could hear it on a speaker. Just what might be expected from an extraterrestrial transmission. According to Drake (1986), "It lasted a few minutes and then disappeared." We were "dumbfounded. Could it be this easy? … We were so surprised and unprepared for it, we didn't know what to do…. We didn't have the presence of mind to steer the telescope off the source, which of course is what we should have done." They then set up a separate receiver with a horn looking at the sky, and when the pulsed signal returned ten days later, they saw it on the antenna pointed toward Epsilon Eridani and also with equal intensity from the reference horn. Almost surely, it was an aircraft radar coming in the antenna sidelobes and not from Epsilon Eridani.

### 4.2 The WOW Signal

The Ohio State radio telescope, known as "The Big Ear"  was a fixed meridian transit parabolic reflector.  When John Kraus (1977) finished surveying the sky for radio galaxies and quasars, there was little left to do in radio astronomy.  For nearly the next 25 years Kraus used the Big Ear to survey the sky looking for signals from an extraterrestrial civilization. To automate the process, the receiver output was digitized and displayed on a simple time vs. frequency plot with limited dynamic range. On 15 August 1977, there was a brief burst of noise covering only a single 10 kHz frequency channel and lasting about a minute, the time taken by a source to pass through the stationary antenna beam. Although it wasn't seen in second antenna beams that swept past the same position a few minutes later, Jerry Ehman impulsively scribbled, "Wow" on the chart record (Ehman 2007,

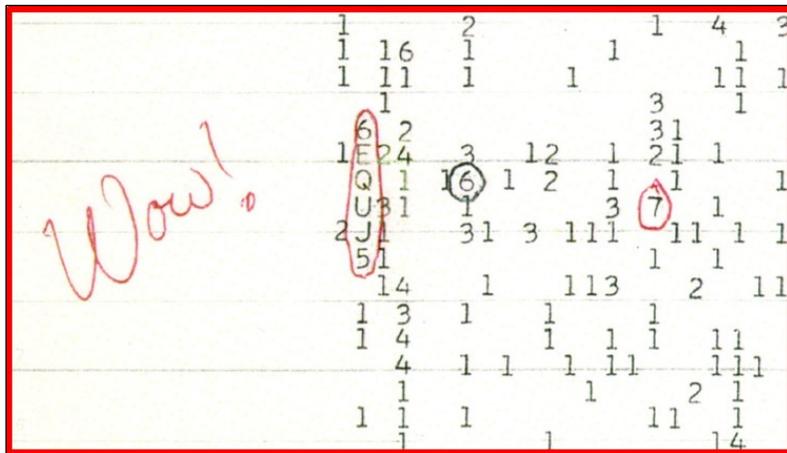

Figure 3.  The famous WOW signal recorded by Jerry Ehman. The x-axis represents a subset of the 50 frequency channels and the y axis and 12 second time samples of the receiver output. Numbers greater than 9 are represented by letters. Taken from Ehman (2007).

Grey 2012). Being a transit instrument, they could not track the source to see if the noise burst would repeat, and it did not appear again on subsequent nights. Most likely it was the kind of interference that radio astronomers deal with all the time, but John Kraus was a good showman, and generated a lot of publicity that inspired numerous follow-up observations, non of which showed any evidence for any repitition of the "WOW" signal.

### 4.3 CTA 102

The radio sources CTA 21 and CTA 102 were the first peaked spectrum sources known (Kellermann et al. 1961). From synchrotron radiation theory, Shklovsky and Kardashev knew that if the peaked spectra were due to synchrotron self-absorption, the sources had to be very small (Slysh 1963, Williams 1963), of the order of 0.01 arcsecond. Along with Parkes 1934 63, Kardashev (1964) considered these peaked spectrum sources as likely candidates for an extraterrestrial transmission. Shkovsky sent his student, Gennady Sholomitky, to Crimea to use a classified military space tracking antenna to study CTA 21 and CTA 102.

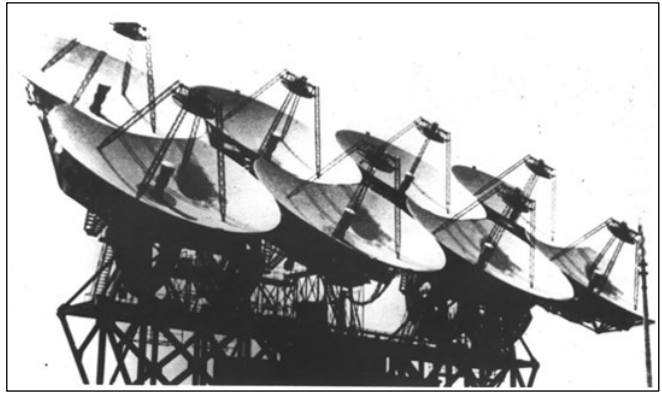

Figure 4. Antenna of the Soviet Deep Space Communication and Control Center in Yevpatoria Crimea, used by Gennady Sholomitsky to observe flux density variations in CTA 102. Space Agency of Ukraine photo.

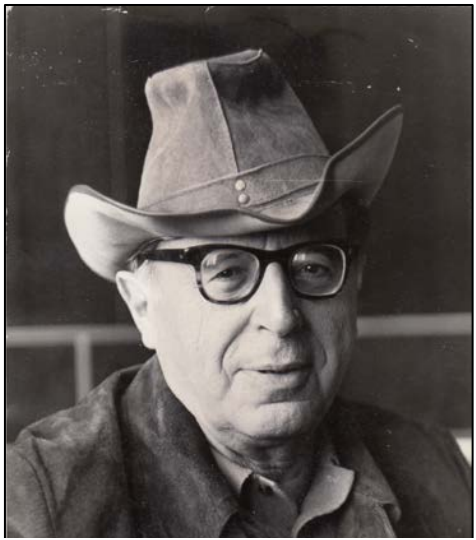

Figure 5. Iosef Shklovsky in 1968 at the 4th Texas Symposium on Relativistic Astrophysics in Dallas, Texas. Photo by the author taken at Shkovsky's request to show his Russian colleagues.

Whether Shklovsky, a strong advocate for SETI, was motivated by his interest in confirming the validity of synchrotron theory, or by detecting evidence for an alien civilization, was known only to Shklovsky.

Sholomitsky (1965) observed both CTA 21 and CTA 102, along with many other radio sources, and reported a remarkable variability of about 30 percent in CTA 102 on a timescale of only a few months. If the flux density of a source varies significantly over a period of say 100 days, then it cannot be more than about 100 light days across; otherwise signals from different parts of the source would arrive at different times and the variations would be smeared out. But, CTA 102 turned out to have a redshift close to one. At this distance, a 100 light day source would be much smaller than 0.01 arcseconds and have a brightness temperature far in excess of the inverse Compton limit so would quickly self-destruct. Sholomitsky, Shklovsky, and Kardashev well understood the problem. On 12 April 1965, a TASS reporter overheard Shklovsky and Kardashev fancifully speculating that the radio emission from CTA 102 might come from an extraterrestrial civilization. The TASS "telegram" reporting on the discovery of Soviet scientists of an artificial cosmic signal motivated Shklovsky and Kardashev to hold a press conference that was attended by Soviet as well as foreign media. When pressed, Kardashev continued to play with the assembled journalists and did not deny the possibility that CTA 102 was an extraterrestrial civilization. The press took it seriously, and the 14 April 1965 edition of *Pravda* reported that extraterrestrials were signaling the Earth. The startling news quickly spread throughout the Soviet Union and around the world, and included two front page stories and an editorial in the 13 April 1965 edition of the *New York Times*.

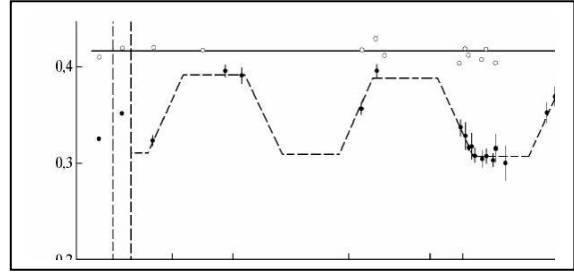

Figure 6. Observed flux density relative to 3C 48 of CTA 21 (open circles) and CTA 102 (filled circles). The plot runs from late 1963 to April 1965. Courtesy of the Sholomitsky family.

When the *Sydney Morning Herald* picked up the report and wanted more information, they contacted the CSIRO Radiophysics Laboratory where I had just completed my study of PKS 1934-63, following up on Kardashev's suggestion that sources like this might be from an extraterrestrial civilization. Being the local "SETI expert," as none of my colleagues wanted anything to do with SETI, the reporter was referred to me. I explained the problem of understanding the rapid variability, and expressed doubts about the SETI interpretation. But when the reporter asked if he could send me his story to check for accuracy, I had to explain that I was leaving in a few days for Moscow. This was a long-planned trip connected with my return to the US from a postdoc in Australia to the US. There was nothing that I could now add to convince the reporter that my trip was to learn about Russian radio astronomy programs and had nothing to do with SETI. The next day, the 18 April 1965 *Sydney Morning Herald*, carried the story.

> SPACE MYSTERY FOR AUST. Scientists pick up signals. Australian Scientists have had mysterious radio signals from an unidentified body, billions of miles away in southern skies, under observation since 1962. This was revealed in Sydney this week following claims by Russian scientists. The Russians claimed that radio signals picked up from a stellar body code-named CTA 102 could be evidence of a super-civilization in outer space. … Dr. Kellermann will fly from Sydney to Moscow on Tuesday to "compare notes" with the Russians.

CTA 102 was further immortalized by the Byrds, a well-known rock group in their memorable recording.[3] We now know that CTA 102 is a typical AGN with a relativistic jet moving toward us at nearly the speed of light, which accounts for the apparent rapid time variations without the need to resort to extraterrestrials.

**4.4 Pulsars**

As is well known, when Jocelyn Bell discovered a pulsating radio source with a 1.33 second repetition rate, Tony Hewish, her dissertation advisor, who was experienced in observational radio astronomy, immediately declared that it was man-made. Bell knew that the pulsating source appeared at the same sidereal time each day, and not at the same solar time and declared that,

"Well, it may be man-made, but not Earth man made." (Bell Burnell 1984). The Cambridge radio group apparently seriously considered whether or not these pulses might have originated from an extraterrestrial civilization, and even fancifully marked their chart recording LGM-1 for Little Green Men. The later suggestion that pulsars were navigational beacons to guide interstellar space craft was given great publicity by the US media.

## 5. SETI Becomes too Important to Leave to the Scientists

Following the publicity surrounding Project Ozma, the highly publicized subsequent false alarms, and a vigorous promotional campaign by a core group of SETI scientists led by Carl Sagan, there was increasing discussion and concern about the moral, legal, ethical, social, technical, medical, and religious impact of a successful detection of signals from an extraterrestrial civilization, one that would be presumably more advanced than our own. Drake's famous 1974 Arecibo transmission to M13 only added to the controversy, with claims that he was irresponsibly disclosing our presence to any alien civilization that might exist on a planet orbiting a star in M13, 21,000 light years away. Apparently, it was time for the US government to step in.

Government involvement in the search for extraterrestrial intelligence started with the summer study organized by NASA and led by Barney Oliver, then the Vice President of Hewlett Packard, and John Billingham, a medical doctor who headed the NASA Ames Committee on Interstellar Communication  The goal of Project Cyclops, as it was known, was to design a radio telescope with at least a factor of 100 improvement in sensitivity over any existing telescope and that would be capable of meaningful search for alien radio transmissions. Oliver's bold vision excited many radio astronomers, but the anticipated cost of more than $10 billion to build up to a thousand 100 meter dishes far exceeded any likely funding, and left a long lasting perception that a realistic search for alien radio signals would be prohibitively expensive. Perhaps not surprisingly, the NASA-sponsored group concluded that NASA should "establish the search for extraterrestrial intelligence as an ongoing part of the total NASA space program with its own funding and budget." The report went on the recommend that "NASA should "establish an office of research and development in techniques for communication with extraterrestrial intelligence," and, "appoint a director and small initial staff." (Oliver and Billingham 1973), p. 171. Billingham now had the blessing he needed to formulate a NASA SETI program. {Fig. 7_Cyclops}

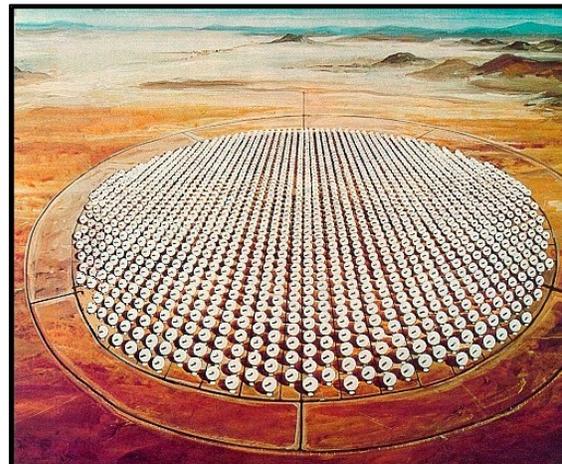

Figure 7. Artist's conception of the thousand element Cyclops array of 100 meter diameter dishes. (Oliver and Billingham 1973)

SETI had apparently become too important, or at least too visible to leave it to the scientists. It appeared that SETI might involve real money, and the politicians began to get involved. In 1975, the US House of Representatives Subcommittee on Space Science and its Applications requested the Library of Congress Science Policy Research Division to report on the "Possibility of Intelligent Life Elsewhere in the Universe." The report compiled by Marcia Smith (1975) was an authoritative account of SETI research primarily in the United States and in the Soviet Union. It served to legitimize a field that most scientists associated with UFOs and science fiction, and was a blueprint for future SETI research. But rather than open the door to new funding opportunities, it may have had the unexpected and unwanted effect of bringing SETI to the attention of Congress.

Although there was no space component, starting with Project Cyclops, the US SETI effort was led for several decades by NASA, with no involvement from the NSF, which normally funds all ground-based astronomy in the US. Encouraged by the House report, NASA initiated a series of Workshops chaired by Phil Morrison.

The Workshop report on *The Search for Extraterrestrial Intelligence* (Morrison, Billingham, and Wolfe 1977) laid out comprehensive plans for the US SETI program for the next decades that followed a set of four bold guidelines.

1. It is both timely and feasible to begin a serious search for extraterrestrial intelligence.
2. A significant SETI program with substantial potential secondary benefits can be undertaken with only modest resources.
3. Large systems of great capability can be built if needed.
4. SETI is intrinsically an international endeavor in which the United States can take a lead.

Meanwhile, not to be outdone, following the 1971 US-USSR meeting in Byurakan, Soviet scientists met again in Gorky in 1972 to formulate their equally bold Research Program on the Problems of Communication with Extraterrestrial Signals that was approved in 1974 by the Board of the Scientific Council on Radio Astronomy of the Soviet Academy of Sciences. The proposed Soviet program included searching nearby stars and galaxies, all-sky surveys over a wide range of radio and infrared wavelengths, finding evidence of Dyson spheres and Bracewell probes, as well as an ambitious program to build dedicated instruments both on the ground and in space to search for signals from extraterrestrial civilizations (CETI 1974). However, aside from a few small investigations, there is no evidence that any of these ambitious programs were actually implemented.

In 1982, the International Astronomical Union established a new Commission on Bioastronomy and SETI, thus recognizing SETI as a legitimate branch of astronomy, and not in the realm of science fiction or UFO studies. In the US, SETI research received a boost and endorsement of legitimacy from the National Academy of Sciences decadal reviews of astronomy (Greenstein 1972, Field 1982, Bahcall 1991, McGee and Taylor 2001). But the NAS reports stressed the need for a variety of approaches including increased support for individually led investigations, and cautioned against large expensive agency-led programs.

With the 1975 Congressional report, the endorsement of the IAU and the NAS, NASA was ready to proceed with an ambitious SETI program (see e.g., Dick 1993). However, in 1978, before the planned program could get off the ground, Senator William Proxmire from Wisconsin awarded his infamous Golden Fleece Award to the NASA SETI program claiming that NASA was "riding the wave of popular enthusiasm for *Star Wars* and *Close Encounters of the Third Kind*," and suggested that SETI should be "postponed for a few million light years."[4] Three years later, Proxmire introduced an amendment to the FY 1982 NASA bill that eliminated any funding for SETI. Nevertheless, the NASA group was able to maintain a low level program under the Congressional radar, and with some quiet diplomacy from Carl Sagan, Proxmire ultimately dropped his opposition to SETI.

NASA convened a second SETI Science Working Group under the leadership pf Frank Drake. The Working Group report (Drake, Wolfe, and Seeger 1984) presented a 14-point set of Conclusions and Recommendations confirming that the most logical approach to SETI was the microwave radio spectrum, that "NASA should remain at the focal point for a well-structured SETI program," and "the search for extraterrestrial intelligence be supported and continued at a modest level as a long-term NASA research program." The recommended modest level long-term program developed into the proposed ten year $100 million *Microwave Observing Project* (MOP) that included both target searches as well as an all-sky survey program, largely disregarding the NAS cautions against large expensive programs. To address the growing tensions between the NASA Ames and the NASA JPL SETI groups, who had proposed a Targeted Search and a less sensitive but more inclusive Sky Survey program respectfully, the MOP included the Solomon-like solution containing both the Ames Targeted Search and the JPL all Sky Survey programs. While the arguments in support of each program were couched in terms of optimum search strategy, the competition for project leadership did not go unnoticed.

With no apparent Congressional opponents (Drake and Sobel 1992) NASA developed their plan for the MOP that would be managed much in the manner as other NASA missions. Much of the next years was spent in developing the needed instrumentation in a field where the technology was developing so rapidly that by time new instrumentation was designed, built, and tested, it was already obsolete. There was competition between Ames and JPL to develop the signal processing hardware known as the *Multichannel Spectrum Analyzer* (MCSA) at Ames and the *Wide Band Spectrum Analyzer* at JPL (see Dick 1993 for detailed history). The MOP, later renamed the *High Resolution Microwave Survey* (HRMS), became a formal NASA mission in 1990 with two project offices at JPL and Ames, a staff of more than 60 people, and a budget of $108 million.

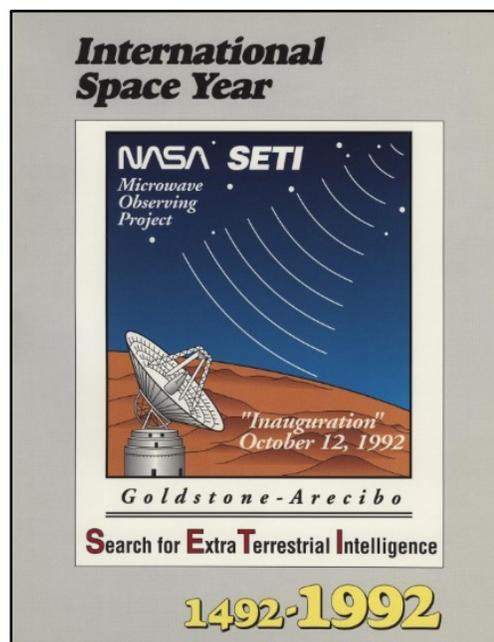

Figure 8. Cover of brochure distributed to the Columbus Day 1992 participants of the joint Arecibo-Goldstone dedication of the Microwave Observing Project.

Although a lot of money had been spent building and testing instrumentation, there were few if any actual observing programs. However, by 1992, NASA was ready to begin their ambitious observing program with a planned simultaneous inauguration at Goldstone and at the Arecibo Observatory that was held on Columbus Day, 12 October 1992, the 500$^{th}$ anniversary of Columbus's landing in the Bahama Islands. At this time, Columbus was still a national hero, although a few decades later he was vilified for tyrannizing the native population. The breakdown in the terrestrial communication link between Goldstone and Arecibo engineers who were hoping to communicate with alien civilizations was perhaps an omen of the serious setback that followed. (Fig 8_MOP)

Normally a $10 million budget item in the $7 billion NASA budget would receive little scrutiny in Congress. Perhaps as a result of intense lobbying and the publicity generated by SETI leaders, the HRMS program came to the attention of Senator Richard Bryan from Nevada. Following an unsuccessful attempt to kill the program in the FY 1992 NASA budget, on 22 September 1993, Bryan issued a press release on *The Great Martian Chase*, claiming,[5]

> As of today millions have been spent and we have yet to bag a single little green fellow. Not a single Martian has said take me to your leader, and not a single flying saucer has applied for FAA approval.

Although a few Senators from both parties spoke in favor of the SETI program, two days later Bryan's late amendment to remove SETI from the NASA budget was passed by a voice vote in the Senate, effectively killing the NASA SETI program. At the time, NASA was still struggling with the embarrassment over the Hubble mirror as well as defending the controversial and expensive Space Station, and wasn't going to make waves over this small SETI budget item that was not part of its mainstream program. An unanticipated consequence of the Senate action was that the NSF, also concerned about its broader program of supporting US science and not offending Congress, for the next decade also quietly refused to fund any proposal for SETI research. Nevertheless, even during this period of NSF withdrawal, SETI programs were quietly, although not secretly, pursued on the NSF-funded Green Bank 140 and 300 foot radio telescopes.

With little prospect for US government funding, the SETI Institute, which was established by Drake and Sagan, successfully initiated "Project Phoenix" that rose from the ashes of the NASA HRMS. Exploiting the instrumentation that had been developed for the HRMS, Phoenix used the Green Bank, Arecibo, Parkes, and Nançay radio telescopes to investigate 800 nearby stars for signs of intelligent life. Meanwhile, the traditional JPL Sky Survey essentially died for lack of funding. With support from the Planetary Society and other private donors, SERENDIP (Search for Extraterrestrial Radio Emissions from Nearby Developed Intelligent Populations) started modestly at the University California Hat Creek Observatory, and expanded to successfully run piggy-back for three decades with ever more sophisticated instrumentation on the world's largest radio telescopes. `SETI@home`, organized by Dan Wertheimer at the University of California, Berkeley, used the screen-saver mode on millions of PCs in some 200 countries to search for signs of intelligence in data from radio telescope, and was, perhaps, the most successful citizen science program ever established. Project META, (Megachannel ExtraTerrestrial Assay) led by Harvard's Paul Horowitz, surveyed the sky for six years around the 21 cm hydrogen line with very high spectral resolution. While finding what appeared to be a small number statistically significant detections concentrated along the galactic plane, none were confirmed by follow-up observations (Horowitz and Sagan 1993).

In 1997, the SETI Institute undertook a new planning study, and this time, unlike the NASA led studies of the 1970s and 1980s, the study was organized independent of any government involvement and included both engineers with a strong digital data processing background as well as SETI enthusiasts. Led by Ron Ekers, a respected Australian radio astronomer who later became President of the IAU, the study report, *SETI 2020,* reviewed the entire field and laid ambitious plans for continued SETI research (Ekers et al. 2002). Interestingly, starting with Project Cyclops, SETI has been a strong driver for increased radio telescope sensitivity. Similarly, the 1971 US-USSR Byurakan meeting recommended building a "decimeter radio telescope with an effective area $\geq 1$ km$^2$" (Sagan 1974), p. 356. One of the interesting outcomes of *SETI 2020* was a design concept for a radio telescope with a collecting area of a million square meters that would impact the planning for the Square Kilometre Array by the international radio astronomy community with Ekers, then Tarter later serving as the first two chairs of the International SKA Steering Committee.

*SETI 2020* recommended a more modest short term goal of building an SKA prototype with a collecting area of one Hectare ($10^4$ m$^2$), dubbed the 1hT. Paul Allen, co-founder of Microsoft, generously contributed more than $30 million to the design and construction of the 1hT, later named the Allen Telescope Array (ATA), the first major facility devoted to SETI. Unfortunately, the lengthy period of design and prototyping used up most of the available funds, so only some 42 of the planned 350 antenna elements were constructed. More recently, the Russian billionaire, Yuri Milner has funded the ten year $100 million *Breakthrough Listen* project, which is conducting the most extensive and most sensitive searches using essentially all of the world's most powerful radio telescopes. Although, like all the other SETI programs, *Breakthrough Listen* has not detected any signs of extraterrestrial intelligence, it has been enormously successful in raising broader awareness of SETI and bringing a new generation of students and scientists to the field.

## 6. Looking Ahead

As Jill Tarter has often noted, it is hard to get funding for continued research on a project that has not had any success even after 60 years and hundreds of observations. She points out that we are looking for the proverbial needle-in-a-haystack, in this case the "cosmic-haystack" with nine degrees of freedom (frequency, modulation, sensitivity, 2 polarizations, 3 directions, and time) (Tarter 2010). At best we have only surveyed a small fraction of phase space, which Tarter has compared to using a glass of water to understand the content of all the water in the Earth's oceans. Probably there is no other area of human inquiry where we know so little about *how* to look, *where* to look, *what* we are looking for, or even *if* there is anything out there to look for.

Six decades of SETI research have perhaps raised more questions than given us answers.

Why, until recently, has SETI been of interest primarily in the United States and in the Former Soviet Union? What was special about these two super powers? Were Soviet and American scientists more concerned about our place in the Universe and the lifetime of intelligent societies than scientists from other countries?

What is the best technique to detect evidence of extraterrestrial intelligent civilizations? Is the detection of artificially transmitted electromagnetic signals or subatomic particles the correct approach? Will the search for evidence of alien technological activity (technosignatures) such as Dyson spheres, or the search for biological activity (biosignatures) such as the detection of methane in stellar atmospheres be more successful? Or, should we be scanning the solar system for Bracewell Probes as the first evidence of extraterrestrial intelligence?

Is the traditional strategy of looking at nearby stars (under the lamppost) the correct approach? Should we be looking nearby for Kardashev Type I civilizations, or for Type II and III civilizations in more distant stars or other galaxies? The answer depends on the SETI luminosity function. If there are a relatively large number of powerful sources, the strongest signals may come from distant stars or from other galaxies. On the other hand, if the SETI luminosity function is steep, the conventional strategy of looking at nearby stars will be most likely to lead to success. In this respect, it may be relevant to recall that brightest radio sources in the sky are distant radio galaxies or quasars, while nearby, so-called, "normal" galaxies are only weak radio emitters.

What is the search optimum strategy? Large focused multi-person national or international teams, or small individual efforts? Where should the money go, to big programs or grants to individual researchers?

What is the right balance between government and private funding? Historically, in almost every country, but especially in the US and the USSR, SETI research was supported either directly or indirectly through government funds. The privately funded Planetary Society, the SETI Institute with its many donors, Paul Allen, and Yuri Milner changed that. In the likely case that Breakthrough Listen finds no extraterrestrials after ten years and $100 million, will there be an incentive (funding?) for further SETI research?

Will the first detection come from a dedicated SETI investigation or will it be from some conventional astronomical research program? How will we distinguish between new physics from the work of an advanced extraterrestrial civilization? If past history is any guide, the answer will be "new physics." Or at least new phenomena. Will a radio amateur be the first to hear and decipher the first extraterrestrial signals, and not a radio astronomer?

What are the implications of success? What will be the scientific, ethical, social, moral, political, legal and religious impact to society? Will it be all good? What are the dangers? How should we react to a successful SETI detection? Obvious, first confirm; keep it quiet until you are absolutely sure it isn't interference, a hoax, or a newly discovered natural phenomena. What then? Whom do you trust? Do you tell no one and continue to observe and try to decipher the signals? Do you tell your government? For Americans, is it the President, the Secretary of State, or the Chairman of the Joint Chiefs of Staff? Since this isn't a national issue do you instead contact the UN? Or do you hold a press conference and tell everyone in the world? In 2010, A Committee of the International Academy of Astronautics issued a *Declaration of Principles Concerning the Conduct of the Search for Extraterrestrial Intelligence,* urging public disclosure of any confirmed detection, including the UN and the IAU, and no transmission of a reply without international consultation.[6] Noticeably, national governments were not included in the protocol. It is uncertain, however, how an individual researcher will behave in a real life situation. Will they follow the protocol, or will they follow their own self interests?

What are the implications of continued failure? Are we alone? Will failure to contact any intelligent extraterrestrial civilization imply that the quantity, *L,* in the Drake Equation is small? If so is this the result of global war, careless overuse of resources, global pandemic, an asteroid impact or some other global catastrophe that we haven't yet thought of? Even cognizant of Morrison and Coconni's proclamation that, "the probability of success is difficult to estimate; but if we never search the chances of success are zero," at what point will continued searches with existing technologies be fruitless?

Should we transmit? If no one transmits, no one will receive. Is there a danger in transmitting? What if the first contact is from a Bracewell probe, which would mitigate the communication time gap resulting from the round trip propagation time to distant stars? Most SETI investigators, as well as the general public, probably accept that alien civilizations may not be humanoid, but implicitly assume that they are biological. Is it more likely the first alien signals will come from a machine, possibly a Bracewell probe with advanced Artificial Intelligence located within our Solar System? What if the transmission at the terrestrial end is also a machine and not a human? Will the extraterrestrial probe and the terrestrial counterpart have sufficiently mature Artificial Intelligence to develop a common language and exchange information? Will humans be involved? How far will Artificial Intelligence evolve beyond the level of Alexa, Cortina, Siri, Watson, and self-drive cars, all of which are developments of only the last decades? Will humans still exist when the first contact is made, and if they do, will the terrestrial machines share what they learn from their extraterrestrial counterparts with their human ancestors?

**7. Acknowledgements**

Parts of this paper are based on Chapter 5 of *Open Skies: The National Radio Astronomy Observatory and its Impact on US Radio Astronomy*, 2020, K.I. Kellermann, E.N. Bouton, and S.S. Brandt, Cham, Springer. I am indebted to Paul Horowitz Ellen Bouton and Ron Enders for help in improving the presentation.

# 8. References

Ambartsumian, V.A. 1972. First Soviet-American Conference on Communication with Extraterrestrial Intelligence (CETI). Icarus, 16, 412-414

Bahcall, J. ed. 1991. The Decade of Discovery in Astronomy and Astrophysics: Report of the Radio Astronomy Panel. Washington, National Academy Press, p. 62

Bell Burnell, J. 1984. The Discovery of Pulsars. In Kellermann, K.I and Sheets, B. (eds) Serendipitous Discoveries in Radio Astronomy, Green Bank, National Radio Astronomy Observatory. pp. 160-170

Bracewell, R.N. 1960. Communication from Superior Galactic Communities. Nature, 186, 670-671

CETI 1974. The CETI Program. Soviet Astronomy, 18, 69-75. Russian original. Astromicheski Zhurnal, 51, 1125-1132

Cocconi, G. and Morrison, P. 1959. Searching for Interstellar Communications. Nature, 184, 844-846.

Dick, S.J. 1993. The Search for Extraterrestrial Intelligence and the NASA HRMS: Historical Perspectives. Space Science Reviews. 64, 93-139

Drake, F.D. 1961. Project Ozma. Physics Today, 14, 40-46

Drake, F.D. 1979. A Reminiscence of Project Ozma. Comic Search, 1, (1), 10-21

Drake, F. 1986. The Search for Extraterrestrial Intelligence. Kellermann, K.I. and Seielstad, G.A. (eds). Proceedings of the NRAO Workshop on The Search for Extraterrestrial Intelligence, Green Bank, National Radio Astronomy Observatory.

Drake, F.D. and Sobel, D. 1992. Is Anyone Out There? The Scientific Search for Extraterrestrial Intelligence. New York, Delacorte Press. pp 195-196

Drake, F.D., Wolfe, J.H., and Seeger, C.L. (eds). 1984, SETI Science Working Group Report, NASA Technical Paper 2244. Washington, NASA.

Dyson, F. 1960. Search for Artificial Stellar Sources of Infrared Radiation, Science. 131, 1667-1668

Ehman, J. 2007. The Big Ear Wow! Signal. http://www.bigear.org/Wow30th/wow30th.htm

Ekers, R.D. et al. eds. 2002. SETI 2020: A Roadmap for the Search for Extraterrestrial Intelligence. Mountain View, SETI Press

Field, G. ed. 1982. Astronomy and Astrophysics for the 1980s. Washington, National Academy Press. pp. 150-151

Gindilis, L.M. and Gurvits, L.I. 2019. SETI in Russia, USSR and the Post-Soviet Space: a Century of Research. Acta Astronautica, 162, 64-74

Gindilis, L.M. et al. 1979. Search for Signals from Extraterrestrial Civilizations by the Method of Synchronous Dispersion Reception. Acta Astronautica, 6, 95-104

Grey, R.H. 2012. The Elusive Wow: Searching for Extraterrestrial Intelligence. Chicago, Palmer Square Press

Greenstein, J.L. ed. 1972. Astronomy and Astrophysics for the 1970s, Vol. 1. Washington, National Academy of Sciences

Harris, M.J. 1986. On the Detectability of Antimatter Propulsion Spacecraft. Astrophysics and Space Science, 123, 297-303

Hippke, M. 2021. Searching for Interstellar Quantum Communications. Astronomical Journal, 162, 1-14

Howard, A.W. et al. 2004. Search for Nanosecond Optical Pulses from Nearby Solar-Type Stars. Astrophysical Journal, 613, 1270-1284

Horowitz, P. and Sagan, C. 1993. Five Years of Project META: an All-Sky Narrow-Band Radio Search for Extraterrestrial Signals. Astrophysical Journal, 415, 218-235

Kardashev, N.S. 1964. Transmission of Information by Extraterrestrial Civilizations. Soviet Astronomy-AJ, 8, 217-221. Russian original, Astronomicheski Zhurnal, 41, 282-287

Kellermann, K.I. 1966. The Radio Source 1934-63. Australian Journal of Physics, 19, 195-207

Kellermann, K.I. et al, 1962. A Correlation Between the Spectra of Non-Thermal Radio Sources and their Brightness Temperature. Nature. 195, 692-693

Kellermann, K.I. and Seielstad, G.A. eds. 1986. The Search for Extraterrestriall Intelligence. Green Bank, NRAO

Kraus, J. 1977. The Ohio Sky Survey and Other Radio Surveys. Vistas in Astronomy, 445-474.

Lingam, M. and Loeb, A. 2017. Fast Radio Bursts from Extragalactic Light Sails. Astrophysical Journal Letters, 837, L23-L27

Loeb, A. 2021. Extraterrestrial: The First Sign of Intelligent Life Beyond Earth, London. John Murray.

McKee, R. and Taylor, J. eds. 2001. Astronomy and Astrophysics for the New Millennium. Washington, National Academy Press. Pp 131-132

Morrison, P., Billingham, J., and Wolfe, J. eds. 1977. The Search for Extraterrestrial Intelligence. NASA SP-419. Washington: NASA

Oliver, B.M. and Billingham, J. (eds). 1973. revised, Project Cyclops: A Design Study for a System for Detecting Extraterrestrial Intelligent Life, NASA CR114445. Originally published 1972, revised 1973, reprinted 1996 by the SETI League and the SETI Institute with additional material.

Pasachoff, J.M. and Kutner, M.L. 1979. Neutrinos for Interstellar Communication. Cosmic Search, 1, (3), 2-8

Sagan, C. (ed.) 1973. Communication with Extraterrestrial Intelligence (CETI). Cambridge, MIT Press

Schwartz, R.N. and Towns, C.H. 1961. Interstellar and Interplanetary Communication by Optical Masers. Nature, 190, 205-208

Sholomitskii, G.B., 1965. Fluctuations in the 32.5-cm Flux of CTA 102. Soviet Astronomy, AJ, 9, 516. English translation from Astronomicheski Zhurnal, **42**, 673

Shostak, G.S. 1993. Third Decennial US-USSR Conference on SETI. San Francisco, Astronomical Society of the Pacific

Slysh, V.I., 1963. Angular Size of Radio Stars. Nature, 199, 682

Sullivan, W.T. III . 1982. SETI Conference at Tallinn. Sky and Telescope, 63, 350-353

Smith, M. 1975. Possibility of Intelligent Life in the Universe. Report prepared for the Committee on Science and Technology, US House of Representatives, Ninety-Fifth Congress, (Washington: Government Printing Office), updated in 1977 to include new astrometric information and the status of the NASA SETI program.

Tarter, J.C. 1980. A High Sensitivity Search for Extraterrestrial Intelligence at λ18 cm. Icarus, 42, 136-144

Tarter, J.C. et al. 2010. SETI Turns 50: Five Decades of Progress in the Search for Extraterrestrial Intelligence. Proceedings of SPIE 7819, Instruments, Methods, and Missions for Astrobiology XIII, 1-13-25

Tovmasyan, G.M. ed. 1964. Extraterrestrial Civilizations. English translation for NASA and the NSF from the Israel Program for Scientific Translation, no. 1823. Springfield, VA, US Dept. of Commerce from Russian original, Yerevan, Armenian Academy of Sciences Press.

Townes, C. 1983. At What Wavelengths Should We Search for Signals from Extraterrestrial Intelligence?, Publications of the National Academy of Sciences, 80, 1147-11

Troitsky, V.S. et al. 1973. Search of Sporadic Radio Emission from Space at Centimeter and Decimeter Wavelengths. Radiofizica, 16, 323-341 (In Russian). English translation in Radiophysics and Quantum Electronics, 16, 239-252

Troitsky, V.S. et al. 1979. Search for Radio Emissions from Extraterrestrial Civilizations. Acta Astronautica, 6, 81

Verschuur, G. 1973. A Search for Narrow Band 21-cm Wavelength Signals from ten Nearby Stars. Icarus, 19, 329-340

Webb, S. 2015. If the Universe is Teeming with Aliens …Where is Everybody? Cham, Springer

Weinreb, S. et al. 1963. Radio Observations of OH in the Interstellar Medium. Nature, 200, 829-831

Williams, P.J.S. 1963. Nature, Absorption in Radio Sources of High Brightness Temperature. Nature, 200, 56-57

Wolszczan, A. and Frail, D. 1992, A Planetary System Around the Millisecond Pulsar PSR1257+12. Nature, 355, 135-147

Zuckerman, B. and Tarter, J. 1980. Microwave Searches in the U.S.A. and Canada,. In Papagiannis, M. D. (ed), Strategies for the Search for Life in the Universe. Dordrecht, Reidel, pp. 81-92

## 9. About the Author

Kenneth Kellermann is an Emeritus Senior Scientist at the US National Radio Astronomy Observatory in Charlottesville, Virginia. His research has been primarily devoted to extragalactic radio sources, especially their radio spectra, time variability, and small scale structure, with short excursions ranging from planetary radio astronomy to cosmology and SETI. After receiving his PhD from Caltech, he spent two years in Australia before joining NRAO in 1965, where he remained for the next 55 years, with short sabbatical interludes at Caltech and CSIRO, as well as a two year stint as a Director at the Max Planck Institute for Radio Astronomy in Bonn, Germany. He is a member of the US National Academy of Sciences, the American Academy of Arts and Science, the American Philosophical Society, and a Foreign Member of the Russian Academy of Sciences. He is a recipient of the 1971 Helen B. Warner Prize, 1973 Rumford Medal, the 2014 Catherine Bruce Gold Medal. Kellermann was a member of the two NASA SETI Science Workshops as well as the later SETI Institute SETI 2020 studies. He has organized and participated in a number of SETI conferences, including the 1971 Byurakan CETI conference.

Figure Captions

Figure 1. Barney Oliver, Hewlett Packard Vice President for Research, early SETI pioneer, originator of the "water-hole" concept, and leader of the 1971 Project Cyclops summer study.

Figure 2. Frank Drake and the author taken in July 2019 during an astrobiology conference. The setting is the lounge where Drake first presented his 1961 Drake equation, shown on the memorial plaque in the background.

Figure 3. The famous WOW signal recorded by Jerry Ehman. The x-axis represents a subset of the 50 frequency channels and the y-axis 12-second time samples of the receiver output. Numbers greater than 9 are represented by letters. Taken from Ehman (2007).

Figure 4. Antenna of the Soviet Deep Space Communication and Control Center in Yevpatoria Crimea, used by Gennady Sholomitsky to observe flux density variations in CTA 102. Space Agency of Ukraine photo.

Figure 5. Iosef Shklovsky in 1968 at the 4$^{th}$ Texas Symposium on Relativistic Astrophysics in Dallas, Texas. Photo by the author taken at Shkovsky's request to show his Russian colleagues.

Figure 6. Observed flux density relative to 3C 48 of CTA 21 (open circles) and CTA 102 (filled circles). The plot runs from late 1963 to April 1965. Courtesy of the Sholomitsky family.

Figure 7. Artist's conception of the thousand element Cyclops array of 100 meter diameter dishes. (Oliver and Billingham 1973)

Figure 8. Cover of brochure distributed to the Columbus Day 1992 participants of the joint Arecibo-Goldstone dedication of the Microwave Observing Project.

Notes

[1] List of Radio SETI Searches, https://technosearch.seti.org/
[2] See https://library.nrao.edu/public/misc/videos/seti1.html for video recordings of the SETI@50 workshop
[3] https://www.youtube.com/watch?v=OONsT-z1hc8
[4] Senator William Proxmire, Press Release, 16 February 1978
[5] Senator Richard Bryan, Press Release, 22 September 1993
[6] https://www.seti.org/protocols-eti-signal-detection